\documentclass[
]{ceurart}

\usepackage{listings}
\begin{document}

\copyrightyear{2024}
\copyrightclause{Copyright for this paper by its authors.
  Use permitted under Creative Commons License Attribution 4.0
  International (CC BY 4.0).}

\conference{Building a Metaverse for All: Opportunities and Challenges for Future Inclusive and Accessible Virtual Environments, May 11, 2024, Honolulu, Hawaii.}

\title{Generative AI for Accessible and Inclusive Extended Reality}


\author[1]{Jens Grubert}[%
orcid=0000-0002-3858-2961,
email=jens.grubert@hs-coburg.de,
url=https://www.mixedrealitylab.de/,
]

\address[1]{Coburg University of Applied Sciences and Arts}

\author[2]{Junlong Chen}[%
orcid=0000-0002-7375-6525,
email=jc2375@cam.ac.uk
]
\address[2]{Department of Engineering, University of Cambridge}

\author[3]{Per Ola Kristensson}[%
orcid=0000-0002-7139-871X,
email=pok21@cam.ac.uk
]
\address[3]{Department of Engineering, University of Cambridge}

\begin{abstract}
  Artificial Intelligence-Generated Content (AIGC) has the potential to transform how people build and interact with virtual environments. Within this paper, we discuss potential benefits but also challenges that AIGC has for the creation of inclusive and accessible virtual environments. Specifically, we touch upon the decreased need for 3D modeling expertise, benefits of symbolic-only as well as multimodal input, 3D content editing, and 3D model accessibility as well as foundation model-specific challenges.
\end{abstract}



\maketitle

\section{Introduction}

Artificial Intelligence-Generated Content (AIGC) is impacting a growing number of individuals, professions, whole societies, and political systems. AIGC already allows for high-fidelity text \cite{bubeck2023sparks}, image \cite{rombach2022high}, and video generation \cite{singer2022make, openaisora2024} alongside further modalities (see recent survey \cite{foo2023ai}). It has been projected to have a substantial impact on future XR (and Metaverse) experiences \cite{chamola2023beyond}. Text-to-3D generation \cite{li2023generative} is particularly relevant for the generation of content for Extended Reality (XR). We foresee that AIGC in general, and the creation and editing of 3D content in particular can have a substantial impact on accessible and inclusive XR environments. Within this paper, we discuss selected promises and potential challenges when using AIGC for creating and interacting with such environments.

\section{Promises of AIGC for Inclusive and Accessible XR Environments}

\subsection{No 3D Modeling Expertise Necessary}
Tools for immersive content generation and editing inside VR became popular in recent years (such as Gravity Sketch\footnote{https://www.gravitysketch.com/ last access February 27th, 2024}) and complement more traditional desktop-based 3D computer graphics tools. Still, such tools typically require a substantial amount of expertise for the generation of 3D models, which are of particular relevance for XR environments. AIGC-based tools promise to make 3D modeling easier to use through simple text-to-3D generation. While initial methods \cite{poole2022dreamfusion, lin2023magic3D} tended to have heavy computational demands requiring hours to generate individual objects, substantial improvements have been made both in terms of computational needs \cite{liu2023one, sun20233D} and visual fidelity \cite{long2023wonder3D, tang2023volumediffusion, wu2024hd} with further advancements being published regularly. Integration of such tools directly inside an immersive environment, as recently proposed by Weid et al. \cite{weid2024gendeck}, could lower the entry barrier for immersive content generation, and, through this, contribute to more accessible 3D content creation for virtual environments.

\subsection{Symbolic Input as an Effective Interaction Paradigm}
Related, the reliance on text as the main input for the creation of 3D assets potentially facilitates accessible (3D) content generation for user groups with limited manual dexterity. Even people with limited speech abilities could potentially create 3D content through gaze-based text entry (\cite{lu2020exploration}) and then navigate and interact within the virtual environments using gaze (c.f., \cite{prithul2022evaluation}). This could contribute to the broader agenda of accessible XR by Design \cite{mott2019accessible, creed2023inclusive}. Still, to allow for accessible XR for a wide range of users (e.g., also one with involuntary eye or limb movements \cite{creed2023inclusive}), multiple, ideally equivalent input modalities should be offered.

\subsection{Using Multimodal Input to Reduce Interaction Complexity}

Inspired by early work multimodal input \cite{bolt1980put, cohen1998efficiency}, we foresee chances that, using a combination of speech and pointing gestures, an AI-based system could support the efficient selection (and manipulation) of objects, that would otherwise require tedious individual interactions. Imagine a user standing inside a virtual or augmented living room that should be re-decorated. For example, now, the user could recolor all objects in the room by pointing and stating "make all objects with a texture like this [the user is pointing towards an object] brighter". Similarly, referencing and understanding real-world objects in augmented reality environments can now already be facilitated through scanning of the physical environment,  spatial representation (e.g., as point cloud), and subsequent processing through large multimodal or large language model \cite{hong20243d, wang20243d}. Through this modifications or add-ons to existing physical objects could be achieved. For example, a user could point towards a window and ask the system to generate a suitable curtain. Similarly, a user could point towards a wall and ask the system to propose artwork that fits the room the user is in. 


Further, multimodal input methods enabled by AIGC also facilitate accessibility design when certain input modalities pose challenges to the user (c.f., \cite{creed2023inclusive}). For example, object recognition and scene interpretation algorithms can convey scene information to the user via speech in lieu of visuals, while augmentative and alternative communication (AAC) techniques based on large AIGC models allow non-speaking individuals with motor disabilities to communicate \cite{shen2022kwickchat}.


\section{Challenges of AIGC for Inclusive and Accessible XR Environments}

\subsection{3D Content Editing and 3D Model Accessibility}
While there is a growing number of tools for the initial creation of 3D assets, subsequent editing of the created 3D assets still often relies on traditional workflows, hence, limiting its applicability for accessible XR content generation. A few works already allow for subsequent editing of initially created 3D scenes but typically rely on specific spatial scene representations such as neural radiance fields (NERFS) or using Gaussian splatting (GS), e.g., \cite{dong2024vica, haque2023instruct, zhuang2024tip}. Challenges remain in also enabling text-guided editing for traditional 3D object representations such as meshes or integrating NERF/GS scene representations in traditional rendering pipelines.  

Further, even today, manually created 3D content poses accessibility challenges through the lack of 3D content metadata \cite{mott2019accessible}, which might be even further complicated through automated 3D asset creation.  This metadata is of importance, for example, to allow for alternative rendering methods such as haptic or auditory rendering for people with visual impairments. Hence, in the future, there should be a focus on how to utilize, e.g., large language models, in the creation process to generate meaningful metadata alongside the actual 3D representation.  Otherwise, in the future, AIGC without metadata for alternative rendering methods could jeopardize carefully designed accessible VR or AR environments. For example, partially empty spaces or holes (e.g., a generated chair or table) in an otherwise accessible virtual room could degrade the user experience.


\subsection{Foundation Models}
While foundation models in AI are a key driver for the recent success of AI in general and content generation in particular, they come with their own set of challenges. 

For one, bias in foundation models is an insufficiently addressed challenge for text and multimodal foundation models \cite{gichoya2023ai, manvi2024large}. It remains an open question how to balance the urge for inclusive 2D and 3D content generated content without replicating (e.g., gender or racial) stereotypes \cite{bianchi2023easily, nozza2022measuring} with concerns about the reliability of generated content regarding (e.g., historical) facts. Recent media attention to Google's Gemini model generating biased depiction of humans (e.g., putting people of color in Nazi-Era uniforms \cite{nyt2024gemini}) exemplifies this challenge. It seems plausible that similar issues can arise when generating 3D content as common text-to-3D generators rely indeed on text-to-image generation first, followed by an uplifting process. But even, when using text-to-3D diffusion directly (e.g., \cite{tang2023volumediffusion}) bias in the underlying 3D model databases such as Objaverse-XL \cite{deitke2024objaverse}, remains challenging, specifically when considering using such content for the creation of inclusive XR environments

Foundation models also have the risk of hallucinating information \cite{rawte2023survey}. The model is encapsulated in a black box, and users have limited agency and knowledge of the computation inside. Models often do not explicitly inform the user of the level of uncertainty in their response. 

Apart from the bias and hallucination problem which can lead to discrimination, exclusion, and toxicity, there are numerous other social and ethical risks of using foundation models in general and large language models in particular (c.f., \cite{weidinger2021ethical} for an overview). For example, in XR environments, pairing a chatbot with a visual (3D) avatar (or other representation) could amplify the risk of over-reliance by anthropomorphizing the system (c.f., \cite{weidinger2021ethical}). 

Related to these discussions is the question of whether or not to rely on closed-source models from large corporations such as Meta or OpenAI or to facilitate the development of open-source models for AIGC-supported XR environments. Both approaches have their own set of challenges such as lack of transparency for closed-source systems to the ease of creating malicious applications using open-source models \cite{chan2023hazards, henderson2023self} which again could inhibit the creation of inclusive XR content.

\section{Conclusion and Future Work}
 Within this paper, we looked at how AI-based content generation might benefit but also challenge the creation of and interaction in inclusive and accessible virtual environments. While some benefits, such as easier and more accessible content generation seem graspable, the potential risks of employing AIGC-based systems should not be underestimated. Some of those risks (such as bias) relate to underlying issues of using foundational models in the first place. However, it is important to be aware of how those problems could manifest themselves in inclusive and accessible virtual environments in the future. As one of our next steps, we plan to focus on supporting efficient content selection and manipulation to facilitate accessible XR environments. We will also explore the possibility of incorporating multimodal manipulation techniques to counterbalance the negative effects of AIGC-based techniques with the accuracy and precision advantages of traditional input techniques.


\bibliography{sample-ceur}


\end{document}